\documentclass[aps,prl,twocolumn,showpacs,groupedaddress,amsmath,amssymb]{revtex4}

\usepackage[dvips]{graphicx}

\begin{document}

\title{Slowing and stopping of chemical waves in a narrowing canal}

\author{Hiroyuki Kitahata}
 \affiliation{Department of Physics, Graduate School of Science, Kyoto University, Kyoto 606-8502, Japan}

\author{Ryoichi Aihara}
 \affiliation{Frontier Research System, RIKEN, Wako, Saitama 351-0198, Japan}

\author{Yoshihito Mori}
 \affiliation{Department of Chemistry, Ochanomizu University, Bunkyo-ku, Tokyo 112-8610, Japan}

\author{Kenichi Yoshikawa}
 \email[To whom correspondence should be addressed. Tel:+81-75-753-3812. Fax:+81-75-753-3779. Email:]{yoshikaw@scphys.kyoto-u.ac.jp}
 \affiliation{Department of Physics, Graduate School of Science, Kyoto University, Kyoto 606-8502, Japan}
 
\begin{abstract}

The propagation of a chemical wave in a narrow, cone-shaped glass capillary was investigated. When a chemical wave propagates from the wider end to the narrower end, it slows, stops, and then disappears. A phenomenological model that considers the surface effect of the glass is proposed, and this model reproduces the experimental trends.

\end{abstract}

\date{\today}

\maketitle

\section{Introduction}

The mechanism of information processing in neuronal systems is one of the central issues in modern science, including biology, biochemistry, and physics. It is known that information processing in nerve cells involves a temporal change in the membrane potential. It has been well-established that, in cablelike nerve cells, pulses on the membrane propagate at a constant velocity and amplitude \cite{nervous_system, nervous_system2}. Recently, on the basis of advances in experimental techniques, it is becoming clearer that the manner of this pulse propagation is critically dependent on the thickness or width of the nerve cells. For example, it has been reported that the dendritic shape of nerve cells strongly affects the propagation of an excited wave \cite{dendrite, dendrite2, nerve_size2}. These effects enable information processing embedded in time-dependent signals, such as coincidence detection, in which the timing of a pair of inputs can be determined \cite{nerve_size, motoike, motoike2}.

On the other hand, it has been claimed that the essence of the neuronal system can be understood in terms of a simple differential equation, the FitzHugh-Nagumo equation, as a type of reaction-diffusion system \cite{fitzhugh, nagumo}. Thus, a neuronal pulse is a manifestation of a dissipative structure generated in nonequilibrium open systems \cite{prigogine}. Several experimental and theoretical studies have been undertaken on the reaction-diffusion systems as a model of the neuronal system.

In such studies, the Belousov-Zhabotinsky (BZ) reaction is often used as an experimental model of the reaction-diffusion systems \cite{bz,bz2}. In this system, chemical waves propagate at a constant velocity and amplitude and, thus, exhibit the characteristics similar to those of pulses in nerve cells. The characteristics of the BZ reaction can be well described using a numerical model, the Oregonator \cite{oregonator,oregonator2}. For example, the Oregonator model reproduces the existence of oscillatory and excitable states, the formation of fascinating patterns (such as target patterns and spiral patterns) \cite{bz2}, and also features associated with hydrodynamical effects \cite{kitahata}. The Oregonator model has a mathematical structure that is similar to the FitzHugh-Nagumo equation. Therefore, to better understand information processing in the neuronal system, it may be useful to compare the behavior of chemical waves in the BZ reaction, although there exists a fundamental difference between them: a neuronal pulse propagates on the two-dimensional tubular membrane and a chemical wave propagates in the three-dimensional volume.

Over the past decade, the behavior of the chemical waves in a narrow space has been studied using a gel system \cite{gel,gel2}, droplet system \cite{muller}, and bead system \cite{aihara,aihara2}. T\'{o}th and co-workers reported that a chemical wave does not propagate when it expands from a thin capillary to the bulk solution. They claimed that a chemical wave can perform information processing \cite{toth,toth2}. Masere et al. performed experiments on the propagation of a chemical wave in a vertical capillary. They discussed their experimental results with special emphasis on the gravitational effect \cite{masere}. In these previous studies, the features of the chemical-wave propagation in a narrow straight capillary with a constant diameter were examined.

\begin{figure}[t]
\begin{center}
\includegraphics{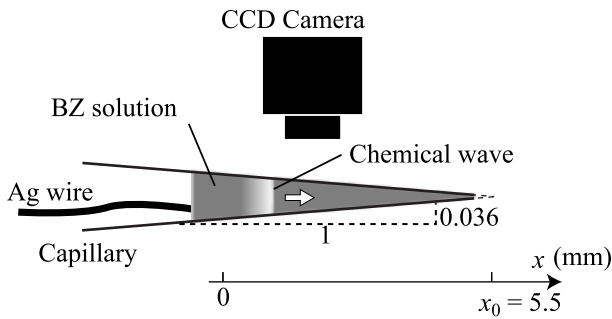}
\end{center}
\caption{Schematic representation of the experimental system; the capillary was arranged horizontally and observed from above by a charge-coupled device (CCD) camera.}
\label{fig1}
\end{figure}

In the present study, we show that a chemical wave slows, stops, and then disappears in a narrowing glass capillary. We discuss this phenomenon in term of the surface volume ratio for the reaction field, i.e., the surface can seriously affect propagation of the chemical wave.

\section{Experiments}

All chemicals were analytical-grade reagents and used without further purification. An aqueous solution of ferroin, or tris (1,10-phenanthroline) iron (II) sulfate, was prepared by mixing stoichiometric amounts of 1,10-phenanthroline and ferrous sulfate in pure water. The water was purified with a Millipore Milli-Q system. The BZ medium contained 0.35 M sodium bromate (NaBrO$_3$), 0.39 M sulfuric acid (H$_2$SO$_4$), 0.12 M malonic acid (CH$_2$(COOH)$_2$), 0.05 M sodium bromide (NaBr), and 4.0 mM ferroin (Fe(phen)$_3^{2+}$) (excitable condition). We allowed the BZ solution to stand for $\sim$ 10 min and then stirred it to diminish the formation of bubbles in the solution.

We prepared a glass capillary (borosilicate glass with a length of 100 mm and inner radius of 0.75 mm; Sutter Instrument Co.) with a decreasing width by pulling a straight capillary with a Micropipete puller (Model P-97/IVF, Sutter Instrument Co.). The both ends of the capillary were open, and the gradient of the inner radius was 0.036. To make it easier to show the experimental results, the $x$-axis is set as shown in Fig. \ref{fig1}. The capillary was immersed in the BZ medium, fulfilled with the BZ medium inside it, and then situated on a horizontal plate for observation with a charge-coupled device (CCD) camera (Digital Microscope Unit, Keyence). As soon as bubbles formed inside the capillary, we stopped the observation: the data reported in the present study are only from the experimental series without bubble formation. All the experiments were performed at room temperature (20 $\pm$ 3 \symbol{23}C). Images were recorded on a videotape and analyzed by an image processing system (Himawari, Library, Inc.).

\begin{figure}[th]
\begin{center}
\includegraphics{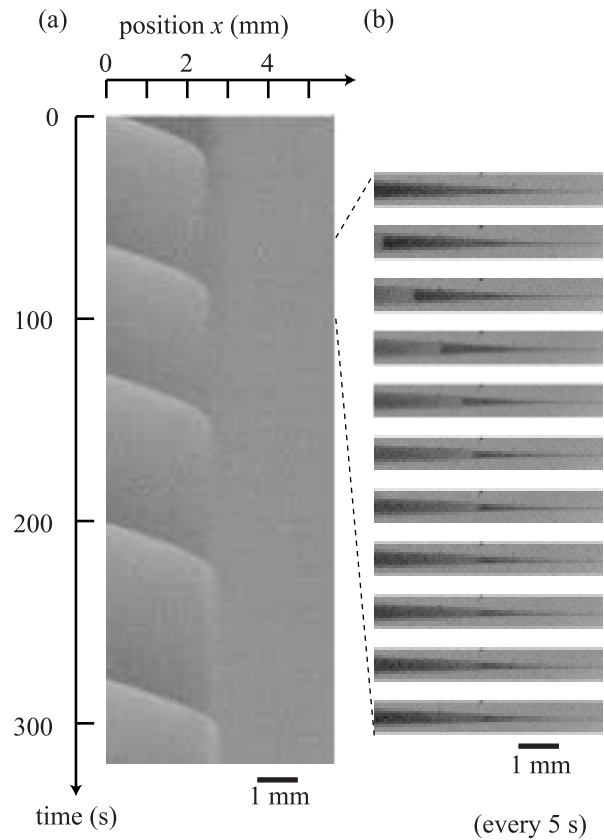}
\end{center}
\caption{Experimental results on the propagation failure in a narrow capillary (the chemical wave propagated from left (wider end) to right (narrower end)): (a) Spatio-temporal plot of chemical-wave propagation and (b) snapshots with an interval of 5 s. When the wave propagated to the narrower region, it slowed, stopped, and then disappeared. Darker and brighter areas correspond to reduced and oxidized states, respectively.}
\label{fig2}
\end{figure}

\section{Results}

\begin{figure}[t]
\begin{center}
\includegraphics[height=11cm]{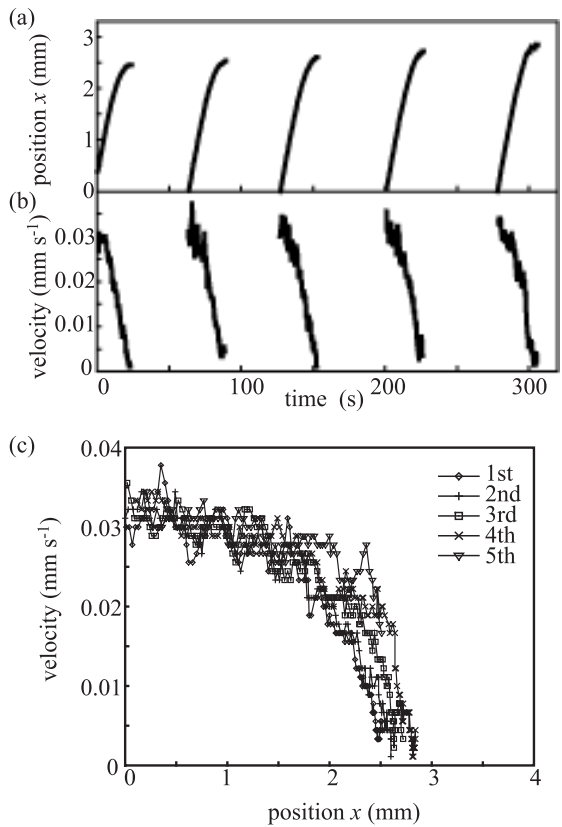}
\end{center}
\caption{Features of the propagation failure of chemical waves in the experiments as shown in Fig. \ref{fig2}: (a)temporal change in the position of the chemical waves, (b)temporal change in the velocity of the chemical waves, and (c)the relationship between position and velocity.}
\label{fig3}
\end{figure}

A chemical wave was initiated at the wider end by a silver wire without disturbance. A chemical wave propagating toward the narrower end slows, stops, and eventually disappears. The oxidized solution in the capillary then reverts to the reduced state. A spatio-temporal plot of images of the capillary along the long axis is shown in Fig. \ref{fig2}(a), and the snapshots (taken every 5 s) are shown in Fig. \ref{fig2}(b). These clearly show that the velocity of the chemical wave gradually decreases before it stops. This feature was observed in every chemical wave initiated. With each subsequent wave, the position at which the chemical wave stops has a tendency to shift slightly toward the narrower end. Figure \ref{fig3} shows the temporal change in the position (Fig. \ref{fig3}(a)) and velocity (Fig. \ref{fig3}(b)) of the chemical wave, and the velocity depending on the position (Fig. \ref{fig3}(c)). From the observation, the point where the chemical wave propagation stopped $x_{\rm s}$ was 2.7 $\pm$ 0.2 mm. We have confirmed the reproducibility of the experimental trend through a number of several repeated experiments.

\section{Discussion}

\begin{figure}[t]
\begin{center}
\includegraphics{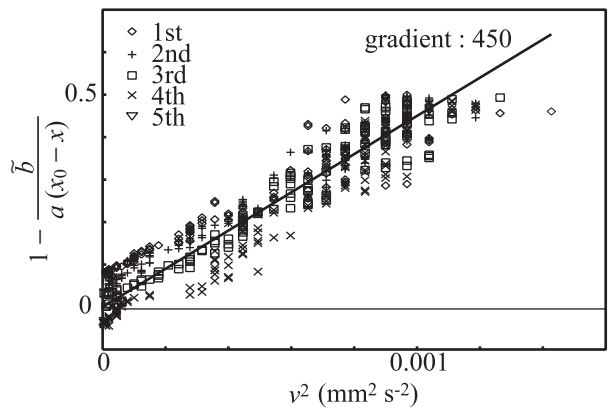}
\end{center}
\caption{Plot of $v^2$ vs $1- \tilde{b}/ [ a ( x_0 - x ) ]$. Five chemical waves were measured, and the different marks correspond to the different waves. By line-fitting, the gradient was calculated to be 450 mm$^{-2}$ s$^2$.}
\label{fig4}
\end{figure}

\begin{figure}[t]
\begin{center}
\includegraphics{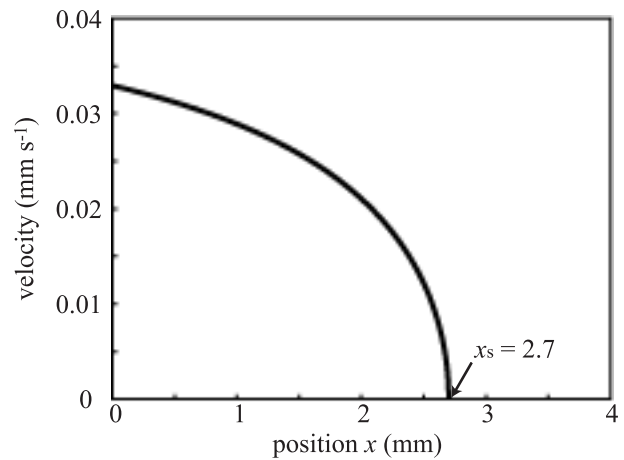}
\end{center}
\caption{Relationship between the position and velocity of a chemical wave, based on the theoretical discussion given in eq.(\ref{eq-v}); the parameters used here are determined from the experiments.}
\label{fig5}
\end{figure}

Here, we discuss here the characteristic change in the manner of wave propagation in a glass capillary. According to an analysis with the Oregonator, the velocity of the chemical wave in the BZ reaction can be written as \cite{tyson-book}
\begin{equation}
v \approx \frac{1}{2} \left( k_5 H A D \right)^{1/2},
\label{velocity}
\end{equation}
where $H$ is the concentration of H$^+$ ions, $A$ the concentration of bromate ions, $D$ the diffusion constant of the activator, and $k_5$ the rate constant of the component reaction, which is introduced as an autocatalytic process by summation of the two elemental processes in the Oregonator model \cite{tyson-book}. Equation (\ref{velocity}) is valid for a plane and steady traveling wave but does not apply to a wave that shows a change in velocity. By assuming that the concentration dependence of the local velocity along the capillary is also given in a similar way, we simply adapt eq.(\ref{velocity}) to interpret the experimental trends in a semiquantitative manner.

The $x$-axis is set in the direction of wave propagation. We describe the inner radius of the glass capillary ($r$) by assuming it to be a part of a cone:
\begin{equation}
r = a ( x_0 - x ),
\end{equation}
where $a$ corresponds to the gradient of the inner radius of the glass capillary and $x_0$ is the position that would be the top of the cone.

The concentration of hydrogen ions ($H$) can be decreased by the effect of the glass surface of the capillary. Note that $H$ is the average concentration over the cross section, and that the effect on the hydrogen concentration by the glass surface is effective only near the surface. Amemiya et al. reported that the effective concentration of hydrogen ions is not homogeneous in the porous glass system \cite{surface}. Thus, the Na$^+$ ions that are connected to the SiO$_2$$^-$ groups could be replaced by H$^+$ ions, which causes a decrease in the value of $H$ near the glass surface. We assume that this effect is proportional to the surface volume ratio:
\begin{equation}
H_{\rm eff} = H \left( 1 - b \frac{2\pi r}{\pi r^2} \right) \equiv H \left( 1- \frac{\tilde{b}}{r} \right),
\end{equation}
where $b$ is a positive constant that reflects the effect of the glass surface. Although, at present, it seems difficult to evaluate $b$ experimentally, as a future subject, it may be valuable to try to measure the change in the chemical environment near the surface. The parameter $\tilde{b}$ ($= 2b$) is a positive constant that is introduced to simplify the equation.

From the abovementioned equations, we can derive the relationship between the position and velocity:
\begin{equation}
v = v_0 \left[ 1- \frac{\tilde{b}}{a ( x_0 - x )} \right]^{1/2},
\label{eq-v}
\end{equation}
where $v_0$ is the velocity of the chemical wave without any effects from the glass surface. By letting $v=0$ in eq.(\ref{eq-v}), we can determine the position $x_{\rm s}$, which is where the chemical wave stops:
\begin{equation}
x_{\rm s} = x_0 - \frac{\tilde{b}}{a}.
\label{eq5}
\end{equation}
Equation (\ref{eq-v}) can also be written as
\begin{equation}
\left( \frac{v}{v_0} \right)^2 = 1- \frac{\tilde{b}}{a ( x_0 - x )}.
\label{eq-v2}
\end{equation}
From the experimental results, the parameters were determined to be $x_{\rm s}$ = 2.7 $\pm$ 0.2 mm, $x_0$ = 5.5 mm, and $a$ = 0.036. Using these values, we plotted the relationship between $v^2$ and $1- \tilde{b}/ \{ a ( x_0 - x ) \}$ in Fig. \ref{fig4}, where $\tilde{b}$ is calculated as $\tilde{b} = 0.20 \pm 0.01 $ using eq. (\ref{eq5}). The results show a linear relationship between these two values. By line-fitting, the gradient was calculated to be 450 $\pm$ 50 mm$^{-2}$ s$^2$. This value corresponds to $1/v_0$$^2$, indicating that $v_0 = 0.047 \pm 0.003 $ mm/s, which roughly corresponds to the wave velocity obtained through the experiment using a wide capillary (0.06 $\pm$ 0.01 mm/s). Using this value, the plot between $v$ and $x$ is shown in Fig. \ref{fig5}. The relationship between $v$ and $x$ in the experiments, as shown in Fig. \ref{fig3}(c), was well-reproduced in a quantitative manner. The velocity profile near the point where a chemical wave stops can be thus explained.

In the abovementioned discussion, the decrease in the effective concentration of the hydrogen ion, ($H_{\rm eff}$), near the glass surface seems to be induced by the adsorption by the SiO$_2$$^-$ groups on the glass surface. With time, this substitution continues, and the effect of the glass surface on $H$ decreases. This may be the cause of the slight shift in the position at which the chemical wave stops.

Because we performed the experiments under an oxygen atmosphere, it is possible that the oxygen disturbs the propagation of the chemical wave. In regard to this problem, we conducted the observation using a straight capillary with open ends and confirmed that absorption of the oxygen from an open end has a negligible effect on the manner of wave propagation. Therefore, it is most probable to expect that the significant factor is the absorption of the H$^+$ ion to the inner wall of the glass capillary.

\section{Conclusion}

When a chemical wave propagates from the wider end of a glass capillary to the narrower end, it slows, stops, and then disappears. This failure to propagate can be caused by the effect of the glass surface inside the capillary. A phenomenological model was considered, and the essential aspects of the slowing and stopping of the chemical wave were reproduced.

This study may promote a further understanding of signal transduction in the neuronal system, especially in narrow tubular parts, such as in the dendrites of nerve cells.

\section*{Acknowledgment}

We thank Dr. J. Gorecki (Polish Academy of Science, Poland), Prof. T. Yamaguchi (National Institute of Advanced Industrial Science and Technology, Japan), and Prof. M. Hara (RIKEN, Japan) for their discussions regarding the mechanism of propagation failure. This work was supported in part by Grants-in-Aid for the 21st Century COE (Center for Diversity and Universality in Physics), Scientific Research B, and JSPS Fellows (No. 5490).

\end{document}